\documentclass{article}

\usepackage{fullpage}

\usepackage{amssymb}
\usepackage{amsthm}
\usepackage{float}
\usepackage{amsmath}
\usepackage{graphicx}
\graphicspath{{figures/}}
\usepackage{float}
\usepackage{booktabs}
\usepackage{setspace}

\newcommand{\fsv}{\ensuremath{\Theta_{vv}}}%
\newcommand{\fsh}{\ensuremath{\Theta_{h}}}%
\newcommand{\fsc}{\ensuremath{\Theta_{c}}}%
\newcommand{\fsl}{\ensuremath{\Theta_{l}}}%

\begin{document}

\title{Prediabetes detection in unconstrained conditions using wearable sensors}


\author{Dimitra Tatli$^{1,2}$, Vasileios Papapanagiotou$^{1,3}$, Aris
  Liakos$^{4}$, Apostolos Tsapas$^{4}$,\\Anastasios Delopoulos$^{1}$}

\date{
  \footnotesize
  $^{1}$Multimedia Understanding Group, Dpt. of Electrical and Computer
  Engineering, Faculty of Engineering, AUTH, Greece\\%
  $^{2}$Embedded Systems Laboratory, Institute of Electrical and Micro
  Engineering, EPFL, Lausanne, Switzerland\\%
  $^{3}$IMPACT research group, Dpt. of Medicine, Huddinge, Karolinska
  Institutet, Sweden\\%
  $^{4}$Diabetes Centre, Second Medical Department at Ippokratio General
  Hospital, Thessaloniki, Greece\\%
}

\maketitle

\begin{abstract}
  Prediabetes is a common health condition that often goes undetected until it
  progresses to type 2 diabetes. Early identification of prediabetes is
  essential for timely intervention and prevention of complications. This
  research explores the feasibility of using wearable continuous glucose
  monitoring along with smartwatches with embedded inertial sensors to collect
  glucose measurements and acceleration signals respectively, for the early
  detection of prediabetes. We propose a methodology based on signal processing
  and machine learning techniques. Two feature sets are extracted from the
  collected signals, based both on a dynamic modeling of the human
  glucose-homeostasis system and on the Glucose curve, inspired by three major
  glucose related blood tests. Features are aggregated per individual using
  bootstrap. Support Vector Machines are used to classify normoglycemic
  vs. prediabetic individuals. We collected data from 22 participants for
  evaluation. The results are highly encouraging, demonstrating high sensitivity
  and precision. This work is a proof of concept, highlighting the potential of
  wearable devices in prediabetes assessment. Future directions involve
  expanding the study to a larger, more diverse population and exploring the
  integration of CGM and smartwatch functionalities into a unified
  device. Automated eating detecting algorithms can also be used.
\end{abstract}

\textbf{Keywords:} Prediabetes, Continuous Glucose Monitors, Signal Processing,
Machine Learning

\section{Introduction}
\label{sec:introduction}

Prediabetes is the precursor stage of type 2 diabetes (T2D) that is
characterized by impaired glucose tolerance and insulin resistance
\cite{preadiabetes-definition}. Glucose homeostasis in humans involves a complex
interplay between insulin and glucagon that regulates blood sugar levels
\cite{Hers1990}. After eating, blood glucose rises, prompting the pancreas to
release insulin, which in turn facilitates glucose uptake by cells for energy or
storage \cite{Rder2016}. In individuals with prediabetes, the body either
becomes resistant to insulin or fails to produce enough, leading to elevated
blood glucose levels. This condition increases the risk of developing T2D and
cardiovascular complications \cite{Mahat2019}.

Despite affecting around $300$ million people globally, approximately $80\%$ of
all cases are undiagnosed, mainly due to lack of awareness \cite{Rooney2023,
  cdcd-prediabetes}. Prediabetes is a medical condition that does not exhibit
any clear symptoms; thus, it is quite challenging to detect. Prompt diagnosis is
crucial, since lifestyle modifications such as weight management, balanced diet,
and regular exercise can prevent or delay the onset of T2D and its complications
\cite{exercise_in_prediabetes}.

Blood tests commonly used to diagnose prediabetes include fasting plasma glucose
(FPG), oral glucose tolerance test (OGTT), and hemoglobin A1c (HbA1c) levels
\cite{blodd_tests}. The FPG test measures glucose levels after overnight
fasting, while OGTT insulin secretory response to a standardized glucose
load. HbA1c reflects average blood sugar levels in the preceding three
months. However, these tests exhibit various limitations such as missed
transient blood sugar spikes, influence from factors like anemia or race on
HbA1c levels, and variations in testing protocols \cite{NationalIDK, Sacks2011,
  NGSP, Nathan2008, race-ethinicity}. Additionally, reliance on single or few
measurements may not capture continuous glucose fluctuations accurately,
necessitating complementary assessments for accurate diagnosis \cite{SACKS,
  Beck2017, Selvin2007, Qian2016}.

The objective of our research is to identify people with prediabetes using
wearable devices, such as smartwatches and Continuous Glucose Monitors
(CGMs). Our approach involves state-of-the-art signal-processing and
machine-learning methods to analyze acceleration signals and glucose
measurements, in order to accurately classify individuals as either
normoglycemic or prediabetic. We propose two novel feature sets: one based on a
dynamic modeling of the human glucose-homeostasis system, $\fsh{}$, and one
based on parameters derived from the glucose curve of the CGM device, $\fsc{}$.

This non-invasive approach offers a practical and effective way of detecting
prediabetes without the inherent limitations of traditional methods. It
leverages CGM sensors, thereby eliminating the need for blood draws and fasting
restrictions. Moreover, the use of CGMs (integrated in a wrist-worn wearable)
ensures more reliable outcomes by enabling the detection of subtle changes in
daily habits and avoiding inaccuracies related to sampling methods. Extensive
research is currently underway to seamlessly integrate CGMs into smartwatches
\cite{Zafar2022, Sakhala, Vaquer2021, Sankhala2022, Sempionatto2021}; this will
enable the application of our proposed method in a more streamlined and less
intrusive approach (since it will not require a dedicated CGM
device). Furthermore, sophisticated algorithms can detect eating patterns by
analyzing accelerometer data \cite{Kyritsis2021, Kyritsis2019}, further
amplifying the effectiveness of our process. The proposed method intends to
increase awareness and enable timely intervention for prediabetic individuals.

\section{Related Work}
\label{sec:related_work}

The study of metabolic syndromes, including both diabetes and prediabetes, has
been an area of intense focus for researchers for many years. Recent progress in
this field has been greatly facilitated by the development and deployment of
wearable, CGM sensors. These sensors play a crucial role in monitoring glucose
levels, by enabling researchers to collect large amounts of data in almost
real-time. Additionally, progress in machine learning and data analysis
algorithms has enhanced researchers' capability to analyze and interpret this
data effectively.

A promising approach to the identification of individuals at risk for diabetes
involves the use of glucose variability (GV) markers: statistical calculations
of CGM measurements. These markers can provide valuable information on an
individual's glucose control status and can offer a comprehensive assessment of
the glycemic fluctuations that occur over time. Acciaroli \textit{et al.}
\cite{Acciaroli2017} propose a method for classifying individuals into normal,
prediabetic, and T2D groups based on a set of $25$ well-established GV markers
(also used in \cite{fabris2016}). Using logistic regression (LR) researchers
achieve high accuracy in identifying healthy subjects ($91.4\%$), however, the
markers are less effective in distinguishing between PD and T2D, with $79.5\%$
accuracy. Another study from Longato \textit{et al.}  \cite{LONGATO2018141}
distinguishes prediabetics from T2D using the $25$ GV indices of
\cite{fabris2016} extended with another $12$ GV markers and combined with
biometric data, achieving an accuracy of $87.1\%$.

The development and application of dynamic models that accurately depict the
human body's glucose homeostasis system is a topic of great interest in the
research community. Due to its complexity, numerous attempts have been made to
study the glucose homeostasis system, primarily for grasping the functioning of
the human body, rather than identifying prediabetic individuals \cite{Mari2020,
  Vicini1997, DeGaetano2008, Cobelli2014}.  L. Van Veen's research
\cite{Kaufman2023, Ng2022} proposes a differential-equations model that proves
promising in distinguishing individuals based on their metabolic syndrome
status. The parameters derived from the model hold the potential to be an
encouraging tool for effective classification of individuals with impaired glucose
homeostasis (IGH), and individuals with effective glucose homeostasis
(EGH). Consequently, it is evident that mathematical models hold great potential
in identifying prediabetic individuals, and further research in this area is
warranted.

The recent study by Bent \textit{et al.} \cite{Bent2021} aims to identify a
biomarker using data collected non-invasively from a smartwatch equipped with
multiple sensors. They gather data on skin temperature, skin conductivity, heart
rate, accelerometer readings, and detailed meal records from $16$ participants
over two weeks, resulting in $69$ features. Through a random-forest
feature-selection model, they determine the most influential factors affecting
glucose levels and predict their approximate values with $84.3\%$ accuracy.

Other research efforts have focused on detecting T2D and prediabetes using
non-CGM sensor data, such as the PIMA dataset containing biometric data, general
blood test results, and medical data from $768$ individuals
\cite{Sehly2020, Naz2020}. Studies using this dataset have applied
machine-learning algorithms like $k$ Nearest Neighbors ($k$-NN), Support Vector
Machine (SVM), and LR, achieving up to $77.21\%$ accuracy. Deep learning
techniques have also been employed in some studies, achieving an impressive
accuracy of $90\%$, but they struggle to extract features that align with human
physiology and lack meaningful interpretation.

\section{Materials and Methods}
\label{sec:materials_and_methods}

This section describes the proposed methodology for creating a classification
system for normoglycemic vs. prediabetic individuals. Specifically, Section
\ref{sec:study_design} describes the study and data-collection processes,
Sections \ref{sec:fsh} and \ref{sec:fsc} describe the novel features we propose
(related to the glucose homeostasis system and to the glucose curve itself
respectively), and Section \ref{sec:classifier_training} describes the process
for training a Support Vector Machine that performs classification of subjects
to prediabetic vs. normoglycemic.

\subsection{Study design}
\label{sec:study_design}

In this study, we aim at evaluating the effectiveness of our proposed method
using a dataset collected in real-world, unconstrained conditions. The dataset
was specifically created for this study since, to the best of our knowledge, no
existing datasets combine data from CGMs and accelerometers, meal information,
and blood tests.

The data collection was conducted in collaboration with the 2nd Medical
Department at the Ippokrateio General Hospital of Thessaloniki and has been
ethically approved by the Committee of Bioethics of the Medical School of
Aristotle University of Thessaloniki on May 10th, 2023 (141/2023). All
participants provided informed consent before participating, and recruitment and
experimental procedures adhered to institutional and international guidelines
for research involving adult human subjects. Clinical data were collected with
REDCap \cite{Harris2008-ph}.

For ground truth, all study participants underwent blood tests prior to data
collection. These tests consist of the FPG test, OGTT, and HbA1c test;
laboratory measurements included FPG, 2-hour glucose values after a 75-g oral
glucose load and HbA1c. These values constitute the ``golden standard'' feature
set, $\fsl{}$. Based on the results, each participant is classified as normal
(normoglycemic) if all three test results were within normal values, otherwise
they are classified as prediabetic.

Subsequently, each participant was provided with a FreeStyle Libre CGM 1st
generation sensor from Abbott and a Ticwatch E2 smartwatch. The CGM device is an
intermittently scanned CGM device (also called flash glucose monitoring, FGM),
and the smartwatch is a commercially-available Android-based (with Wear OS 2)
device. Participants were instructed to constantly wear the devices for $14$
days without altering any habits of their everyday life. Additionally, they were
strongly encouraged to use an Android application that has been specially
developed for this study to record meals and eating activity (i.e., start and
stop timestamp and a photo of the consumed food). This approach enables the
collection of diverse data streams in order to study glucose regulation and
activity patterns in realistic conditions.

In total, we collected data from $23$ people, all of whom underwent blood tests
according to the research protocol and were divided into two categories: with or
without prediabetes. One participant developed T2D during the study, thus their
data were not used in the analysis. Out of the $22$ people who successfully
participated in the study, $10$ ($45\%$) are in the prediabetes group and $12$
($55\%$) exhibit normal glucose tolerance. In total, $13$ men ($59\%$) and $9$
women ($41\%$) participated. Sample statistics are summarized in Table
\ref{tab:stats}. The dataset consists of a total of $305$ days of recordings and
$821$ meals.

% \todo{Liakos: Mipos tha prepei na anaferontai ksexorista oi times gia kathe
%   omoada h/kai na prostethoun kai alles parametroi (vlepe pinaka 1 me
%   dimografika apo ti vivliografiki anafora \#21 apo tin omada tou van
%   Veen). Episis, mipos xreazontai kai kapoioi peraitero pinakes me summary
%   statistics apo ta dedomena tou CGM kai tou accelerometer opos stin
%   proanafertheisa dimosiefsi?}

% \begin{figure}
%   \centering
%   \includegraphics[width=\linewidth]{stard.drawio.pdf}
%   \caption{STARD diagram}
%   \label{fig:stard}
% \end{figure}

\begin{table}
  \centering
  \caption{Baseline characteristics of the recruited participants}
  \label{tab:stats}
  \begin{tabular}{lcc}
    \toprule
    & \textbf{average}& \textbf{standard deviation}\\
    \midrule
    age [$years$]              & $43.32$ & $16.82$ \\
    weight [$kg$]              & $82.13$ & $16.00$ \\
    waist circumference [$cm$] & $93.86$ & $14.75$ \\
    body mass index [$kg/m^2$] & $27.12$ &  $5.03$ \\
    \bottomrule
  \end{tabular}
\end{table}

% \subsection{Data Analysis}
% \label{sec:data_analysis}

% \todo{I am not sure if this subsection is needed} In this section, we describe
% the process and techniques employed to extract features from the dataset, which
% were subsequently used as inputs for machine learning classification
% algorithms. The dataset contains data collected from various sources including
% CGMs, namely the FreeStyle Libre sensor by Abbott, smartwatches capturing
% accelerometer data, bio-metric data from each user, such as age, weight, etc,
% and meal information (starting and ending timestamps).

\subsection{Glucose-homeostasis feature set $\fsh{}$}
\label{sec:fsh}

This feature set is based on our variation of the mathematical model developed
by Van Veen \textit{et al.} \cite{Ng2022}. This particular model is chosen not
only for its simplicity in accurately describing a complex system with a minimal
number of parameters (three) but also because it only requires glucose
measurements. This detail is critical since other models and approaches
\cite{Palumbo2013} often require additional measurements, such as insulin
levels, which can be challenging to obtain, as they typically require blood
draws. Furthermore, this model has been validated against real-world data
involving multiple individuals with promising results, showcasing the
generalization ability of the model.

According to the Van Veen mathematical model, the dynamics of blood glucose
homeostasis are modeled through the following system of coupled differential and
integral equations:
\begin{equation}
  \label{eq:1}
  \dot{e}(t) = - A_{3} - u(t) \cdot \left(e(t) + \bar{e}\right) + F(t)
\end{equation}
\begin{equation}
  \label{eq:2}
  u(t) = A_{1}e(t) + A_{2} \int_{-\infty}^{t}{\lambda e^{-\lambda (t - \tau)} e(\tau) d\tau}
\end{equation}
where $e(t)$ is the excess glucose concentration from some set value $\bar{e}$,
$\bar{e}$ is the glucose baseline which is calculated by averaging the glucose
values of all the local minima from the entire time series, $\dot{e}(t)$ is the
time derivative of $e(t)$, $F(t)$ is the general glucose intake (i.e., when
eating a meal) and consumption (i.e., when being physically active), $u(t)$ is
negative feedback mechanism that moderates blood glucose concentration as it
deviates from its normal level, $\lambda$ is related to the time scale of the delays
in the feedback mechanism, $A_{1}$ is the coefficient of the proportional and
$A_{2}$ the coefficient of the integral part of the closed-loop glucose control,
while $A_{3}$ is the base metabolic rate, assumed to be constant. Three main
components influence glucose deviation from its baseline according to
Eq. (\ref{eq:1}), in particular $F(t)$, $u(t)$, and $A_{3}$.

The control variable $u(t)$ represents all the mechanisms and the collective
effects that return blood glucose to normal levels, and it is modeled using a
proportional-integral strategy (Eq. (\ref{eq:2})). Regardless of glucose
deviation, the controller acts to restore glucose to baseline levels. Feedback
mechanisms differ for positive and negative deviations, involving insulin
excretion and glucose uptake for hyperglycemia, and glucagon release triggering
liver glucose release for hypoglycemia. The feedback function accommodates these
variations, ensuring stability in glucose regulation.

The model is fitted to the data using a custom gradient descend algorithm
developed by Van Veen \cite{Ng2022}, yielding the optimal values for parameters
$A_1$, $A_2$, and $\lambda$. These parameters are considered as features unique to the
physiology of each individual, enabling discrimination between normoglycemic
individuals and individuals with impaired glucose tolerance, such as prediabetes
or diabetes.

Using this model as a foundation, we enhance its efficacy in two ways: by
incorporating additional data, specifically accelerometer readings, and by
changing the fitting process.

The first enhancement involves expanding the term $F(t)$ of Equation \ref{eq:1}
to specifically account for glucose consumption due to vigorous physical
activity as:
\begin{equation}
  \label{eq:expanded_Ft}
  F(t) = \tilde{F}(t)-A_{4} \cdot c(t)
\end{equation}
introducing a new parameter, $A_4$, which represents the rate at which glucose
is metabolized due to exercise, and the term $c(t)$ which is the rate at which
glucose levels diminish in response to vigorous physical activity, calculated
using accelerometer data and participant biometrics. We estimate $c(t)$ as:
\begin{equation}
  \label{eq:4}
  c(t) = R \cdot m(t)
\end{equation}
where $R$ is the Basic Metabolic Rate (BMR) of each individual and $m(t)$ is the
Metabolic Equivalent of Task (MET) of the activity that is performed by the
individual at time $t$.

BMR is well approximated by the Harris–Benedict equation \cite{Bendavid2021,
  Campos2024}:
\begin{equation}
  \label{eq:bmr}
  R =
  \begin{cases}
    66.473 + 13.752 \cdot \text{weight} + 5.003 \cdot \text{height} - 6.755 \cdot \text{age},& \text{if gender} = \text{male}\\
    665.096+9.563 \cdot \text{weight}+1.85 \cdot \text{height} - 4.676 \cdot \text{age},& \text{if gender} = \text{female}
  \end{cases}
\end{equation}

Respectively, $m(t)$ is a way to estimate the amount of energy (or oxygen
consumption) used during physical activity, relative to the amount of energy
expended at rest. It is a unit that quantifies the intensity of physical
activities based on basic metabolic rate \cite{AINSWORTH2011}. MET can be
estimated using activity counts, which can be easily calculated from
acceleration signals using the algorithm of Brondeel \cite{Brondeel2021}. The
MET prediction based on activity counts per minute is given by
\cite{Crouter2006}:

\begin{equation}\label{eq:met}
  m(t) = 
  \begin{cases}
    1.0 & \text{if } a(t) < 50 \\
    1.83 & \text{if } 50 \leq a(t) \leq 350 \\
    1.935 + 0.003002 \cdot a(t) & \text{if } 350 < a(t) < 1200 \\
    2.768 + 0.0006397 \cdot a(t) & \text{if } a(t) \geq 1200
  \end{cases}
\end{equation}
where $a(t)$ is the activity counts per minute at time $t$. We compute activity
counts as the mean value of acceleration (in units of $g \approx 9.80655$
$\text{m/sec}^2$) across $1$-minute epochs on a filtered version of the
acceleration magnitude, using a Butterworth band-pass filter (range from $0.5$
to $1.5$ Hz).

The revised Equation \ref{eq:1} is:
\begin{equation}
  \label{eq:3}
  \dot{e}(t) = - A_{3} - u(t) \cdot \left(e(t) + \bar{e} \right) + \tilde{F}(t) -
  A_{4} \cdot R \cdot m(t)
\end{equation} 

By introducing the new parameter $A_4$ we incorporate additional metrics which
were not included in the original model, leading to more robustness and
efficiency.

The second enhancement involves changing the gradient descent algorithm. In the
original work of Van Veen \cite{Ng2022}, the model is fitted to each extracted peak
separately, with parameters subsequently calculated using the bootstrap mean
(over all the data of each participant) of all peak-derived parameters. Instead,
we opted to run the gradient descent algorithm concurrently for all peaks,
minimizing the global error. While this approach does not ensure perfect fitting
to each peak individually, it yields better results with improved
generalization.

Consequently, the feature vector of a participant for $\fsh{}$ is
\begin{equation}
  \label{eq:theta_h}
  \mathbf{\theta}_{h} = [A_{1}, A_{2}, \lambda, A_{4}]^{T}
\end{equation}
as defined in Equations \ref{eq:3} and \ref{eq:2}, where each of the four
parameters is estimated using the gradient descend algorithm described above
across all data of each participant (independently).

\subsection{Glucose-curve feature set $\fsc{}$}
\label{sec:fsc}

% A5, fpg_mean, fpg_std, fpg_mean_after_8h

This feature set includes features that are based on parameters extracted from
the glucose curve of the CGM device, in combination with information from the
acceleration signals and the meal diaries. Continuous monitoring of glucose
levels combined with food diaries enables the calculation of various parameters
related to glucose fluctuations, including those influenced by meal timing and
physical activity. These parameters are essential for understanding the dynamic
nature of blood glucose regulation and provide valuable insights into metabolic
health. Inspired by the blood tests typically prescribed by healthcare
professionals to detect prediabetes, namely OGTT, FPG, and HbA1c, we derive
features resembling the measured values.

% For instance, we approximate the HbA1c value of a lab test by using the formula
% approved by the American Diabetes Association \cite{Nathan2008}:

% \begin{equation}\label{eq:HbA1c}
%   \hat{HbA1c} = \frac{eAG + 46.7}{28.7}
% \end{equation}

% where $eAG$ (Estimated Average Glucose) is the average glucose value of all days
% of record. The longer the duration of the glucose recording, the more accurate
% the above formula is.

Approximating FPG lab-test results is straight-forward process. By obtaining
meal events, we detect fasting periods lasting longer than $8$ hours and examine
glucose levels during these time intervals. This approach offers several
advantages over traditional medical examinations. Firstly, it eliminates the
inconvenience of mandatory fasting and takes advantage of naturally occurring
fasting periods (e.g., during sleep). Moreover, numerous spontaneous fasting
events typically occur during the monitoring period, contributing to assessments
based on multiple measurements, thereby enhancing objectivity and reducing
bias. Additionally, this method allows for the exploration of various
measurements associated with fasting periods, such as the mean and standard
deviation of glucose levels during fasting.

Specifically, we produce $3$ features from each such interval: the mean glucose
level, $\mu_{\text{G}}$, the standard deviation of glucose level,
$\sigma_{\text{G}}$, and the mean glucose level after discarding the first $8$ hours,
$\mu_{\text{FG}}$. It should be noted that $\mu_{\text{FG}}$ is an approximation of
the FPG test; FPG mandates a minimum of $8$ hours of fasting and then blood
glucose is measured. In our case, we detect naturally occurring fasting periods
of at least $8$ hours and then estimate the blood test result as the average of
the glucose curve until the next eating event starts.

We also produce a $4$-th feature that captures the rate of glucose consumption
after an eating event, normalized by the level of physical activity, since
physical activity also contributes to glucose consumption. This feature, denoted
$\mu_{\text{NPGG}}$, is the normalized postprandial glucose gradient (NPGG), and
is computed on the basis of each eating event. We obtain the interval from the
maximum (peak) of the glucose curve during the eating event, until the next
local minimum. Let $t_{1}$ and $t_{2}$ denote the start and end of such an
internal. NPGG is then computed as:
\begin{equation}
  \label{eq:npgg}
  \mu_{\text{NPGG}} = \frac{1}{t_{2}-t_{1}} \sum_{t=t_{1}}^{t_{2}} \frac{\dot{e}(t)}{\dot{a}(t)}
\end{equation}
where $\dot{a}(t)$ is the time derivative of $a(t)$. Derivatives are estimated
as first-order discrete gradients.

\begin{figure}
  \centering
  \includegraphics[scale=0.8]{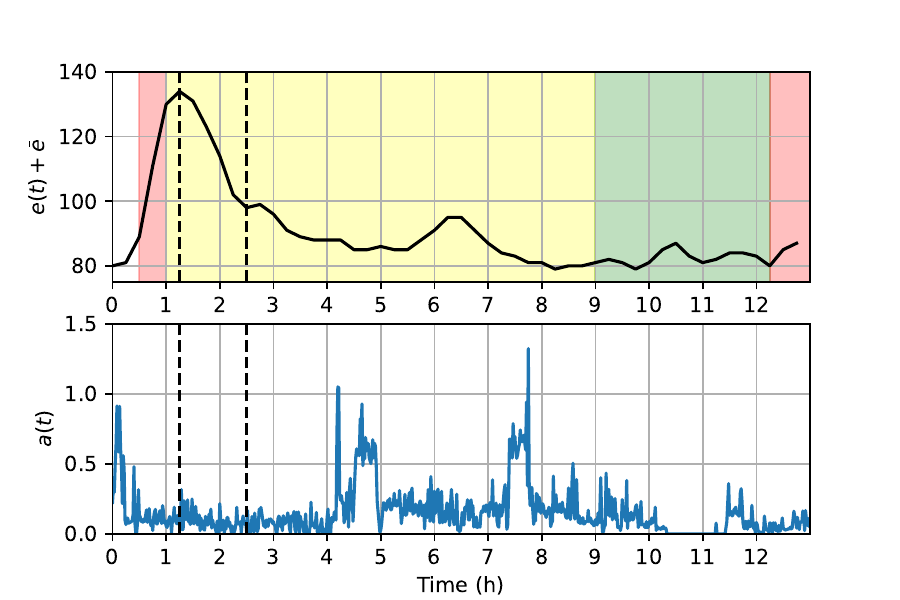}
  \caption{Example of computing the $\fsc{}$ features. Top figure shows the CGM
    curve (black); red areas denote eating, yellow and green denote fasting. The
    yellow area is exactly 8 hours long. Features $\mu_{\text{G}}$ and
    $\sigma_{\text{G}}$ are computed on $e(t) + \bar{e}$ over the fasting (yellow and
    green) interval, while $\mu_{\text{FG}}$ is computed as the average of
    $e(t) + \bar{e}$ over only the green area. NPGG is computed within the
    interval between the dashed black lines (meal peak $t_{1}$ and next local
    minimum $t_{2}$ of Equation \ref{eq:npgg}). Bottom figure shows $a(t)$.}
  \label{fig:data_example}
\end{figure}

Finally, we aggregate multiple occurrences of each feature by computing the mean
value across all valid, detected intervals of a participant, yielding the
feature vector
\begin{equation}
  \label{eq:theta_c}
  \mathbf{\theta}_{c} = [\bar{\mu}_{\text{G}}, \bar{\sigma}_{\text{G}}, \bar{\mu}_{\text{FG}}, \bar{\mu}_{\text{NPGG}}]^{T}
\end{equation}
Figure \ref{fig:data_example} shows an example of collected data (CGM curve and
computed counts, $c(t)$), and which part of the data is used for computing each
feature of $\fsc{}$.

% \begin{figure}
%   \centering
%   \includegraphics[scale=0.8]{lasso_feature_selection}
%   \caption{The features selected most frequently using LASSO}
%   \label{fig:lasso_features}
% \end{figure}

\subsection{Classifier training}
\label{sec:classifier_training}

Extracting the features described in the previous sections yields two feature
vectors for participants, one for each of the two feature sets of $\fsh{}$ and
$\fsc{}$ respectively. In our experiments we evaluate the following three
feature sets: $\fsv{}$ (i.e., the feature set introduced by Van Veet \textit{et
  al.}  \cite{Ng2022}), $\fsh{}$, and $\fsh{}\cup\fsc{}$. We also use $\fsl{}$ as a
baseline feature set for comparison. Note, however, that $\fsl{}$ requires lab
blood tests, while the other three feature sets only require our proposed,
passive data collection system.

We train SVM classifiers on these feature sets, using the radial-basis function
(RBF) kernel. We opt for $C=1$ for the SVM and $\gamma=1/d$ for the RBF kernel, where
$d$ is the number of features (for each feature set), since we standardize our
features. Experiments are carried out in a leave-one-subject-out (LOSO) way
(which is equivalent to leave-one-out in our case since we only have one
feature vector per participant). In addition, we repeat the computation of
performance metrics, in particular accuracy, specificity, sensitivity, and area
under the curve (AUC), for each algorithm $20$ times. The final metric is
determined by taking the mean value over the $20$ iterations; this helps
eliminate any influence from chance-based outcomes.

\section{Results \& Discussion}
\label{sec:results}

\begin{figure}
  \centering
  \includegraphics[scale=0.8]{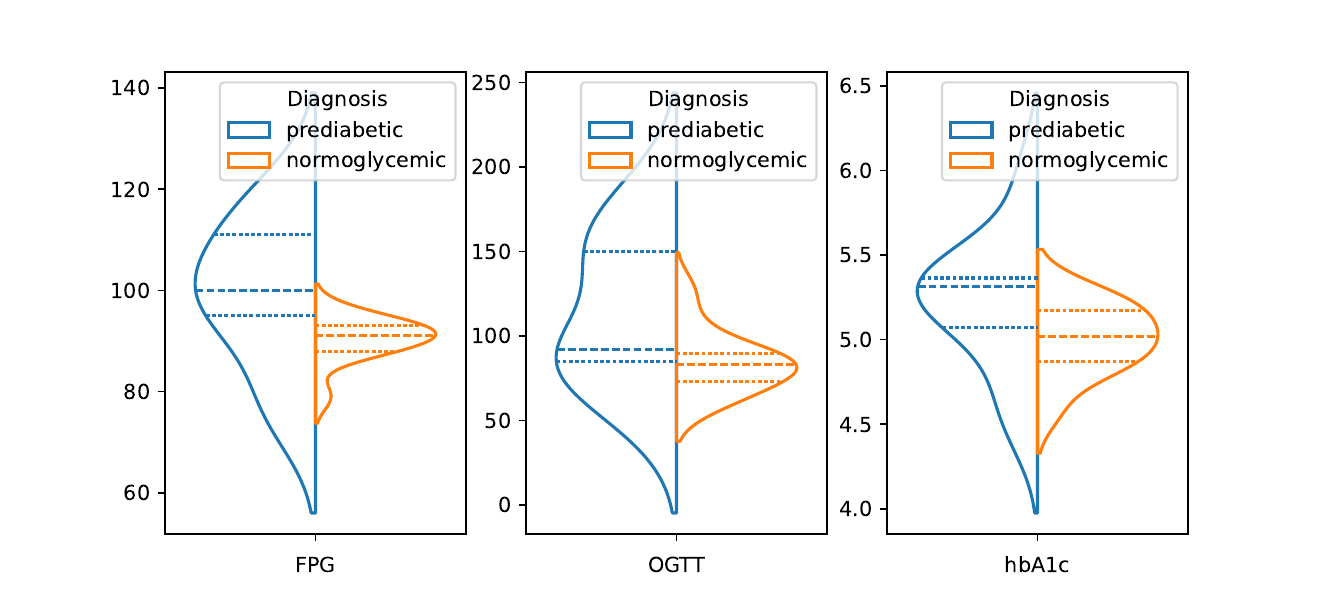}
  \caption{Distribution of each feature of $\fsl{}$ across normoglycemic and
    prediabetic participants}
 \label{fig:lab_features}
\end{figure}

To get an overview of the lab blood test values over the dataset, we present the
distribution of the three tests for normoglycemic and prediabetic participants
in Figure \ref{fig:lab_features}. 

% We also train an SVM as a baseline comparison for our proposed method, and to
% compare against the features of \cite{Ng2022} and the features we propose in
% this work.

\begin{figure}
    \centering
    \includegraphics[scale=0.8]{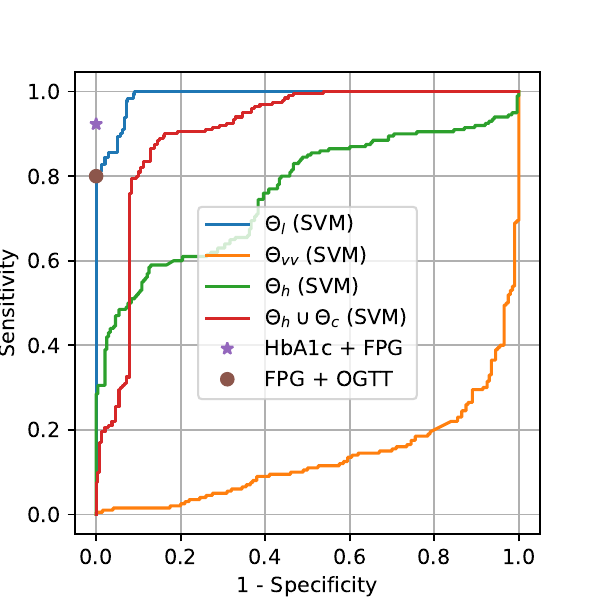}
    \caption{Results of our experimental evaluation. SVM ROC curves for:
      features of lab blood tests $\fsl{}$, $\fsv{}$ features of \cite{Ng2022},
      our version of modified features of \cite{Ng2022} $\fsh{}$, and $\fsh{}$
      features extended with $\fsc{}$. Effectiveness of common medical practice
      of using either the combination of HbA1c and FPG blood tests or the
      combination of FPG and OGTT blood tests.}
    \label{fig:svm-roc-curve}
\end{figure}

SVMs are robust, supervised, machine-learning classifiers; in our experiments,
they exhibit the highest effectiveness with the combined $\fsh{}\cup{}\fsc{}$
feature set (among the three feature sets that do not require lab blood
tests). We compute accuracy, specificity, and sensitivity. We also plot the ROC
curve (Figure \ref{fig:svm-roc-curve}) for each feature set and compute the area
under the curve (AUC). All four metrics are reported in Table \ref{tab:svm}.

Table \ref{tab:svm} shows that our proposed changes to the original model of Van
Veen \textit{et al.} \cite{Ng2022} (i.e., $\fsv{}$ vs. $\fsh{}$) improve
significantly the effectiveness of the classification. Note that the relatively
low overall effectiveness of $\fsv{}$ is not completely unexpected since
parameters $\fsv{}$ and the corresponding model of \cite{Ng2022} have not been
designed for classification of normoglycemic vs. prediabetic individuals
originally, but to model the dynamic glucose homeostasis system of healthy
(i.e., normoglycemic) individuals.

Expanding $\fsh{}$ with feature set \fsc{} improves the classification
effectiveness, increasing accuracy from $0.73$ to $0.86$ and specificity from
$0.5$ to $0.85$, while sensitivity remains at $0.92$. Figure
\ref{fig:svm-roc-curve} shows the ROC curves and the expanded feature set
$\fsh{}\cup\fsc{}$ in general improves over $\fsh{}$. This is also evident from the
increase of AUC from $0.76$ to $0.9$.

We also train an SVM on the three lab tests to get an estimate of what an
automatic classification scheme would attain if the standard blood tests were
used for automatic decision. The SVM trained on $\fsl{}$ achieves the highest
effectiveness overall. However, it should be stated that $\fsl{}$ requires lab
blood tests, while $\fsh{}$ and $\fsc{}$ do not.

We also compute the effectiveness of two approaches commonly used in medical
practice: having only the HbA1c and FPG lab blood tests, and having only the FPG
and OGTT lab tests. In both cases, if any of the two tests exceed well-known
thresholds, the individual is diagnosed as non-normoglycemic. Both of these
approaches achieve superior specificity of $1.0$ compared to the use of SVM and
$\fsh{} \cup \fsc{}$ features which achieves $0.85$. However, our approach with SVM
and $\fsh{} \cup \fsc{}$ achieves sensitivity of $0.92$ which is equal to the one
achieved by HbA1c and FPG and superior to the $0.80$ of FPG and OGTT. This is
encouraging as it shows that our proposed method, which does not require lab
tests, can be used as an effective and efficient screening tool in real
practice.

% \begin{table}
%   \centering
%   \caption{Classification results for the different combinations of classifier
%     types and feature sets.}
%   \label{tab:svm} 
%   \begin{tabular}{llcccc}
%     \toprule
%     \textbf{Classifier} & \textbf{Feature Set} & \textbf{Accuracy}
%     & \textbf{Sensitivity} & \textbf{Specificity} & \textbf{AUC} \\ 
%     \midrule
%     SVM & \fsv{}            & $0.65$ & $0.8$  & $0.5$  & $0.06$ \\
%     SVM & \fsh{}            & $0.73$ & $\mathbf{0.92}$ & $0.5$  & $0.76$ \\
%     SVM & \fsh{} $\cup$ \fsc{} & $\mathbf{0.86}$ & $\mathbf{0.92}$ & $\mathbf{0.85}$ & $\mathbf{0.9}$  \\
%     RF  & \fsv{}            & $0.33$ & $0.25$ & $0.4$  & $0.35$ \\
%     RF  & \fsh{}            & $0.82$ & $0.91$ & $0.7$  & $0.82$ \\
%     RF  & \fsh{} $\cup$ \fsc{} & $0.76$ & $0.83$ & $0.67$ & $0.84$ \\
%     \bottomrule
%   \end{tabular}
% \end{table}

\begin{table}
  \centering
  \caption{Classification results for normoglycemic vs. prediabetic
    participants: SVM with each of the three feature sets of $\fsv{}$, $\fsh{}$,
    and $\fsh{} \cup \fsc{}$, baseline SVM with lab blood tests $\fsl{}$ features,
    and results based on common medical practice where only two lab blood tests
    are used.}
  \label{tab:svm} 
  \begin{tabular}{lcccc}
    \toprule
    \textbf{Method} & \textbf{Accuracy}
    & \textbf{Sensitivity} & \textbf{Specificity} & \textbf{AUC} \\ 
    \midrule
    SVM w. $\fsv{}$            & $0.65$ & $0.8$  & $0.5$  & $0.06$ \\
    SVM w. $\fsh{}$            & $0.73$ & $\mathbf{0.92}$ & $0.5$  & $0.76$ \\
    SVM w. $\fsh{} \cup \fsc{}$   & $\mathbf{0.86}$ & $\mathbf{0.92}$ & $\mathbf{0.85}$ & $\mathbf{0.9}$  \\
    \midrule
    SVM w. $\fsl{}$            & $0.9$  & $1.0$  & $0.78$ & $0.99$ \\
    \midrule
    HbA1c + FPG                & $0.95$ & $0.92$ & $1.00$ & \\
    FPG + OGTT                 & $0.86$ & $0.80$ & $1.00$ & \\
    \bottomrule
  \end{tabular}
\end{table}

% RF classifiers yield very good outcomes as well, but not as great as SVM
% classifiers do. Furthermore, when using feature set \fsh{} only, the RF
% classifier is able to classify individuals with $82\%$ accuracy, $91\%$
% sensitivity, and $70\%$ specificity, i.e., better when compared to using the
% combination of both feature sets \fsh{} and \fsc{}. This highlights the ability
% of the \fsh{} feature set to differentiate between normoglycemic and prediabetic
% individuals; this supports the argument that our changes to the original model
% of \cite{Ng2022} are effective and meaningful.

% \begin{figure}
%     \centering
%     \includegraphics[scale=0.8]{random_forest_results}
%     \caption{Random Forest ROC curve for different feature sets}
%     \label{fig:rf-roc-curve}
% \end{figure}

\section{Conclusions}
\label{sec:conclusions}

In summary, CGM devices and smartwatches are valuable tools for prediabetes
assessment, as they enable prompt identification of dysglycemia while avoiding
the burden of traditional lab blood tests. Our proposed approach demonstrates
promising results in automatic and effective prediabetes detection, achieving
high accuracy and sensitivity which ensures that few undiagnosed cases occur.

However, limitations in data collection, such as a small and biased dataset, and
occasional errors in CGM sensor measurements must be addressed. Future research
efforts should prioritize expanding the dataset to include a more diverse
population and developing integrated devices that combine CGM and smartwatch
functionalities for enhanced data retrieval and assessment efficiency. While our
current study serves as a proof of concept, further research involving a larger
and more diverse population is crucial for definitive conclusions and to fully
understand the potential impact of this approach.

We are optimistic that with additional improvements to the methodology and with
access to more data, our approach can yield even better results, suitable for
clinical applications. The integration of glucose monitoring into smartwatches
would eliminate the need for a separate, stand-alone CGM device, further
increase the deployment capabilities and reducing user intrusiveness. By
integrating our algorithm in a smartwatch-based application, individuals at risk
of prediabetes can be alerted and informed timely. This would facilitate early
lifestyle changes and improve healthcare outcomes.

% \section*{Acknowledgements}

% TODO\todo{Are we citing REBECCA or something here? How is the paper paid?}

% \section*{Statement of Authorship}

% TODO

\section*{Funding sources}

This work was partially supported by the Information Processing Laboratory of
AUTH by undertaking the cost of the CGM and the smartwatch devices.

% \section*{Declaration of Generative AI and AI-assisted technologies in the
%   writing process}

% During the preparation of this work the author(s) used ChatGPT in order to
% effectively communicate their ideas, employ precise terminology and streamline
% their expression to ensure clarity and efficiency. After using this
% tool/service, the author(s) reviewed and edited the content as needed and
% take(s) full responsibility for the content of the publication.



%% References


\bibliographystyle{elsarticle-num}
\bibliography{references.bib}






% \end{thebibliography}

\end{document}